# Diffusion in La$_n$CoIn$_{3n+2}$ phases studied by perturbed angular correlation

Randal Newhouse [a] and Gary S. Collins [b]

Department of Physics and Astronomy, Washington State University, Pullman, WA, USA

[a]randynewhouse@gmail.com, [b]collins@wsu.edu

**Keywords:** diffusion, perturbed angular correlation, PAC, hyperfine interaction, nuclear quadrupole interaction, atomic jump frequency, intermetallic compound.

**Abstract.** Jump frequencies of $^{111}$In/Cd tracer atoms were measured for a series of layered phases La$_n$CoIn$_{3n+2}$ using the technique of perturbed angular correlation of gamma rays (PAC). The frequencies were determined by analysis of nuclear quadrupole relaxation produced by fluctuating electric field gradients. Samples were synthesized having nominal values $n$= 1, 2, 3, 5 and $\infty$, with $n=\infty$ corresponding to the L1$_2$ phase LaIn$_3$. The phases form heuristically from LaIn$_3$ by replacing every $(n+1)^{th}$ (100) mixed plane of La and In atoms with a plane of Co-atoms. For the $n$=1 phase, LaCoIn$_5$, jump frequencies were too small to detect. Two signals were observed, one for indium atoms next to the Co-planes and the other for more distant indium atoms. No relaxation was observed for atoms next to the Co-planes, indicating that there is no diffusion across the Co-planes. With increasing $n$, jump rates for the other In-atoms increased toward values observed for LaIn$_3$. Jump frequency activation enthalpies for $n$= 3 and 5 were observed to be the same as for $n=\infty$, suggesting the same diffusion mechanism. However, the jump-frequency prefactors were found to be smaller for small $n$, which is attributed to reductions in the connectivity of the diffusion sublattice. We conclude that diffusion in the layered phases is remarkably similar to diffusion in LaIn$_3$ once the reduced connectivity is taken into account.

## Introduction

Atom-scale diffusion in solids can be detected through analysis of nuclear quadrupole relaxation that occurs when diffusional jumps lead to changes in magnitude or orientation of electric field gradients (EFGs) at tracer nuclei. Perturbed angular correlation of gamma rays (PAC) is a nuclear hyperfine interaction method that has been applied to study such relaxation [1] using the $^{111}$In/Cd probe. The result of a measurement is a time-domain PAC perturbation function $G_2(t)$ that, in effect, exhibits quadrupole precessions of the 247 keV level of the daughter nuclide $^{111}$Cd. Stochastic fluctuations of EFGs lead to their averaging. Extensive studies have been made of nuclear relaxation in compounds having the L1$_2$ structure using PAC [1, 2, 3, 4]. In that structure, which has A$_3$B stoichiometry, probe atoms jumping on the A-sublattice experience reorientations of the EFG from one [100] cube direction to another in each near-neighbor jump. In the slow-fluctuation regime, when the average jump-frequency of the probe atom, $w$, is less than the quadrupole interaction frequency, motional averaging of the EFGs leads to decoherence of the quadrupole perturbation function and, to a good approximation, the dynamically relaxed perturbation function is given in terms of a static perturbation function $G_2^{static}(t)$ by [5]

$$G_2(t) \cong \exp(-wt) G_2^{static}(t). \qquad (1)$$

Among many L1$_2$ phases studied, the greatest jump frequencies of $^{111}$Cd probes have been observed in In-rich LaIn$_3$ [2], with lower jump frequencies observed in In-poor LaIn$_3$ and in all the other rare-earth tri-indides (RIn$_3$, R= Ce, Pr, Nd,…Lu, Y, Sc) [2, 4].

In the present paper, we examine how diffusivity changes in a confined geometry through measurements on a series of layered phases La$_n$CoIn$_{3n+2}$, whose prototype is Ho$_n$CoGa$_{3n+2}$ [6]. There has been considerable interest recently in phases R$_n$CoIn$_{3n+2}$ because Ce$_n$CoIn$_{3n+2}$ phases have been shown to exhibit heavy-fermion superconductivity [7, 8]. La$_n$CoIn$_{3n+2}$ phases have been studied as non-magnetic analogs of the cerium phases [6]. Crystal structures of $La_nCoIn_{3n+2}$ are shown in Fig. 1 for n=1, 2 and ∞ [6]. The ternary phases are formed heuristically by replacing every $(n+1)^{th}$ (100) mixed plane of La and In atoms in $LaIn_3$ with a plane of Co-atoms, as shown in the figure. The structure approaches the binary $LaIn_3$ structure as *n* increases and the density of Co-planes decreases. Interatomic separations of La and In atoms between the Co-planes remain very similar to separations in $LaIn_3$. The goal of this work is to see how jump frequencies of Cd probe atoms on In-sites change as a function of the spacing index *n*. As will be shown below, Co-planes were found to confine diffusion of atoms on indium sites to layers in between the Co-planes.

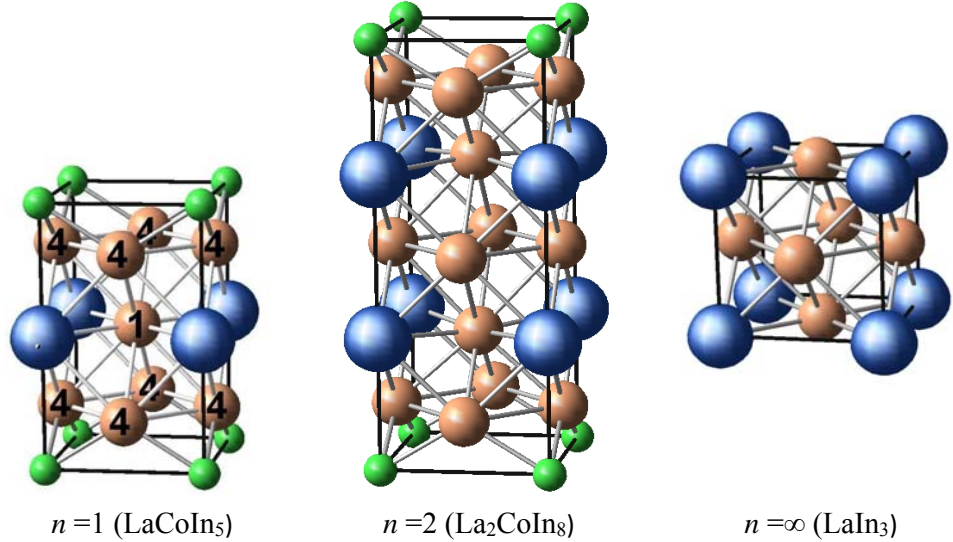

n =1 ($LaCoIn_5$)   n =2 ($La_2CoIn_8$)   n =∞ ($LaIn_3$)

Fig 1. Unit cells of $La_nCoIn_{3n+2}$ for *n*=1,2 and ∞. The n=∞ structure is the cubic $L1_2$ structure, with La atoms at corner sites and In atoms at face-centers. Co atoms are shown at corners of the tetragonal unit cells for *n*=1 and 2. Two inequivalent In-sites in the *n*=1 structure are marked 1 and 4 according to the numbers of atoms per formula unit.

**Sample preparation and measurements**

Samples were prepared by arc-melting high purity metal foils under inert argon gas with $^{111}$In/Cd activity, as in previous work [1,3]. The typical mass of a sample was 100 mg. The ratio of masses of La and Co was carefully controlled in order to produce samples having the appropriate layer index *n*. Since jump frequencies of Cd probes had previously been found to be greater in $LaIn_3$ samples that had an excess of In, samples were made with a slight excess of In-metal over the stoichiometric amount. The mole fraction of $^{111}$In/Cd probes incorporated in each sample was very low, about $10^{-10}$. Samples were annealed at temperatures of about 400°C prior to measurement in order to increase the crystalline order. Signals for all layered phases disappeared due to melting or peritectic reaction at high temperature, with approximate temperatures indicated in Table 1 below.

PAC measurements were made using a four-counter spectrometer of standard design with $BaF_2$ scintillation detectors [9]. Measurements were made at temperatures up to 600°C in a turbopumped vacuum oven. Four coincidence spectra collected at relative angles of 90° and 180° were algebraically combined to obtain the experimental perturbation function $G_2(t)$. For the spin-5/2 PAC excited-state level of $^{111}$Cd, the perturbation function for probe atoms in a random polycrystalline sample experiencing a unique, static, quadrupole interaction is given by

$$G_2^{static}(t) \cong \tfrac{1}{5} + \tfrac{13}{35}\cos\omega_1 t + \tfrac{10}{35}\cos\omega_2 t + \tfrac{5}{35}\cos\omega_3 t , \qquad (2)$$

in which the three frequency harmonics are related via $\omega_3 = \omega_2 + \omega_1$. When there is an axis of four-fold symmetry through the probe site, such as for site-1 in Fig. 1 (left), the three harmonic frequencies have the harmonic relationship $\omega_1:\omega_2:\omega_3 = 1:2:3$ and the EFG is termed axially symmetric. When symmetry is lower, such as for site-4 in Fig. 1, then there is no harmonic relationship, although the ratio $\omega_2/\omega_1$ is constrained to be within the range 1-2; the signal is then termed nonaxially symmetric. In the latter case, the quadrupole interaction is characterized by the fundamental frequency $\omega_1$ and by an EFG asymmetry parameter $\eta$. Experimental PAC perturbation spectra are typically superpositions of perturbation functions for each of the various lattice locations occupied by the probes, and are fitted to determine the site fractions, fundamental frequencies and asymmetry parameters. For further information about PAC spectroscopy, see [9].

**Room temperature measurements and attributions of signals to sites**

Fig. 2 shows PAC spectra for samples of compositions $n= 1,2,3$ at room temperature. For $n=1$ (LaCoIn$_5$, top), the spectrum exhibits two signals: a low frequency, axially symmetric signal that has a fundamental quadrupole interaction frequency of 56 Mrad/s, similar to that observed in LaIn$_3$, and a high-frequency, nonaxially symmetric signal with fundamental frequency 214 Mrad/s. Examining the crystal structure shown for $n=1$ in Fig., 1 (left), it can be seen that there are two inequivalent indium-sites, with one atom at a point of tetragonal local symmetry at the center, and four atoms at equivalent, lower-symmetry points. These are accordingly labeled 1 and 4 in Fig. 1.

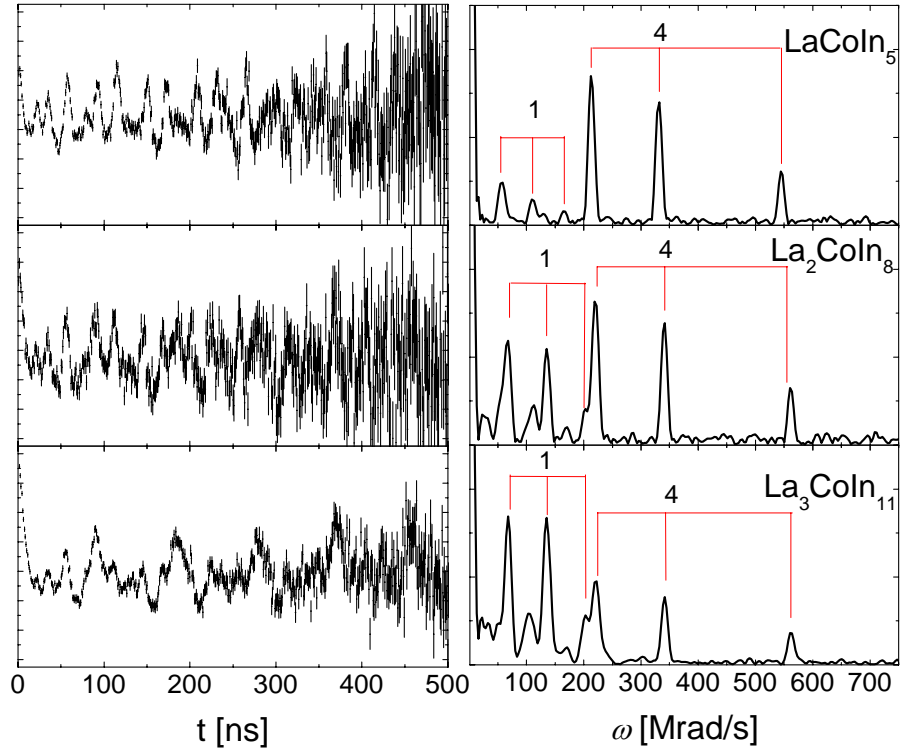

Fig 2. PAC spectra (left) and Fourier transforms (right) of La$_n$CoIn$_{3n+2}$ samples prepared to have mean compositions $n=1,2,3$, measured at room temperature. All three spectra are dominated by two signals, each of which has three component frequency harmonics indicated by the inverted tridents drawn on the Fourier transforms. As explained in the text, the two signals are related to sites marked 1 and 4 in Fig. 1 for the $n=1$ structure.

Thus, for LaCoIn$_5$ ($n=1$), one should observe a 4:1 ratio for the fitted, experimental site fractions, corresponding to the numbers of indium atoms at the two sites. The observed ratio was 3:1, as can be seen from amplitudes of the fundamental frequencies in the Fourier transform of the spectrum for LaCoIn$_5$ (Fig. 2, top right). Thus, the symmetries of the two sites, site fraction ratios, and similarity of the 56 Mrad/s fundamental frequency to the one in LaIn$_3$, clearly identify the two signals with the two indium sites. Accordingly, we attribute the two signals to sites 1 and 4, and label them as such in Fig. 2. Mean quadrupole interaction parameters for $n$=1,2,3,5 and $\infty$ obtained from fits of spectra measured at room temperature are listed in Table 1. For samples prepared to have $n$=2, 3 and 5, principal features of the PAC spectra were the same two signals observed for $n$=1, and we identify them also as sites 1 and 4 in Fig. 2 and in the Table.

Table 1. Quadrupole interaction parameters and ratios of site fractions of the two dominant signals, measured at room temperature.

| $n$ | Phase | $\omega_1$ (site 1) Mrad/s | $\omega_1$ (site 4) Mrad/s | $\eta$ (site 4) | $f_4/f_1$ | $4/(3n-2)$ | Melt or peritectic temp |
|---|---|---|---|---|---|---|---|
| 1 | $LaCoIn_5$ | 56.0(3) | 213.5(1) | 0.49(1) | 3.0(3) | 4 | ~900 K |
| 2 | $La_2CoIn_8$ | 66.7(5) | 218.7(3) | 0.500(2) | 1.04(6) | 1 | ~875 K |
| 3 | $La_3CoIn_{11}$ | 67.5(2) | 222.9(4) | 0.513(3) | 0.39(1) | 0.571 | ~725 K |
| 5 | $La_5CoIn_{17}$ | 65.6(1) | 214.7(6) | 0.497(4) | 0.22(1) | 0.308 | ~725 K |
| $\infty$ | $LaIn_3$ | 67.9(3) | - | - | - | 0 | 1413 K |

Further examination of crystal structures in Fig. 1 helps to justify the assignments of signals to "sites 1" and "site 4" in the Table. It can be seen from Fig. 1 that the structure for $La_2CoIn_8$ ($n$=2) has three inequivalent In-sites: (a) four atoms close to the Co-layers, (b) two atoms in planes with La atoms, and (c) two atoms in a median plane. However, local coordinations of atoms (b) and (c) are both the same as in $LaIn_3$. Since the total EFG at a site is dominated by atoms within the first coordination sphere , they should have similar quadrupole interaction parameters. Consequently, one can anticipate also observing only two signals for $La_2CoIn_8$, one for two atoms at site 4 positions, and the other a composite signal for two atoms with quadrupole interaction parameters similar to that observed in $LaIn_3$. As shown in the table, the observed ratio of site fractions, 1.04(6), is in excellent agreement with the predicted value 1.0. Ratios of the two site fractions obtained from fits of spectra for samples prepared to have $n$=3 and $n$=5 layer indices are in fair agreement with predicted ratios, which in general are given by $f_4/f_1 = 4/(3n-2)$. It should be noted that it was difficult to control the precise ratio of masses of La and Co during sample preparation due to slight mass losses during melting, so that actual compositions may differed from nominal ones. In addition, the stoichiometry of a sample could of course have been intermediate between ideal La:Co ratios, in which case there would have been an irregular distribution of spacings of Co-planes.

**Nuclear relaxation at elevated temperature**

PAC spectra were measured at elevated temperatures for samples having nominal composition parameters $n$= 1,2,3 and 5. Spectra were fitted with two signals according to Eq. 1 in order to obtain relaxation frequencies $w$. No relaxation was detected for the site 4 signals in any samples at any temperature. From this, it is concluded that indium atoms adjacent to the Co-layers do not diffuse appreciably. It should be noted that jumps of indium atoms back and forth between site 4 sites across the Co-planes would be invisible to PAC measurements because the EFG does not change, but that jumps of atoms from site 4 to either site 1 or to other sites 4 in the same plane would produce nuclear relaxation, and can therefore be ruled out. Consequently, long range diffusion of indium atoms across the Co-layers can be considered to be blocked.

Significant relaxation was observed for site 1 signals. For $LaCoIn_5$, jump frequencies at all temperatures were near or below the level of detectability, about 1 MHz. However, significant relaxation was observed for n=2, 3 and 5 that increased with n. To illustrate, Fig. 3 shows spectra for $La_3CoIn_{11}$ at increasing temperatures, showing damping of the ~65 Mrad/s signal of site 1 (100 ns period) while the ~225 Mrad/s signal of site 4 is unaffected by temperature.

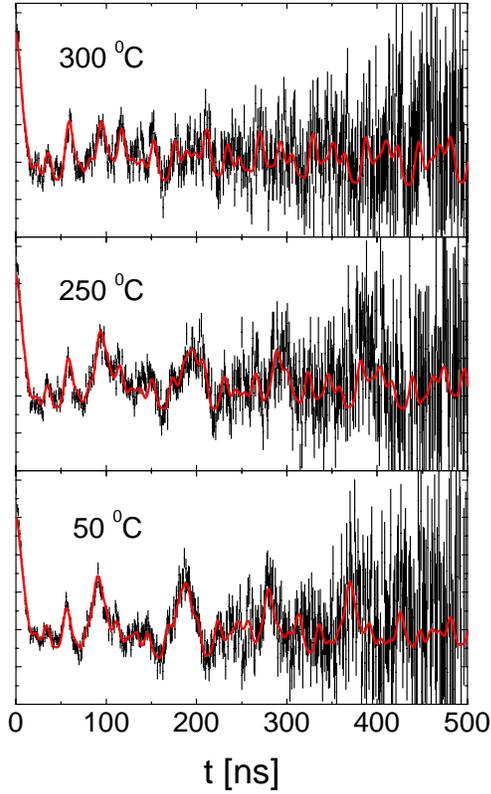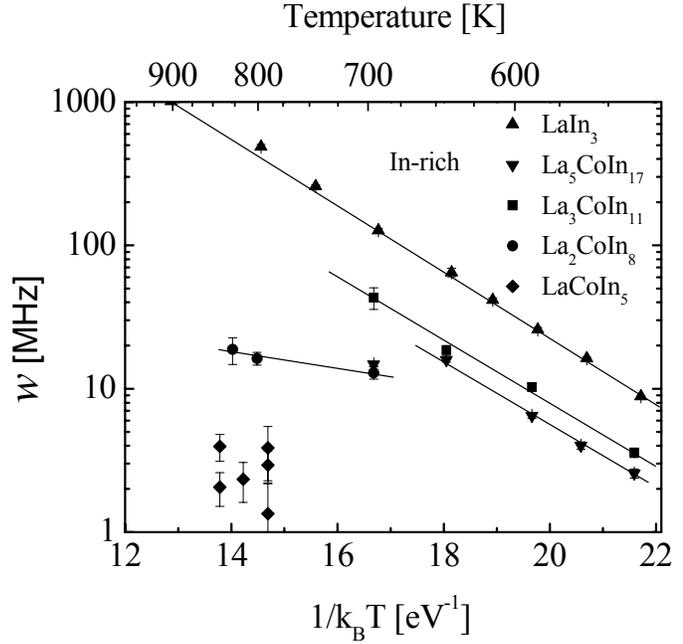

Fig. 3. PAC spectra for $La_3CoIn_{11}$ at elevated temperatures. With increasing temperature, the site-1 signal, with period 100 ns, becomes damped out.

Fig. 4. Arrhenius plot of jump frequencies obtained from fits of nuclear relaxation for sites 1 in the indicated phases.

Fig. 4 shows an Arrhenius plot of jump frequencies for sites-1 in the various phases, including data for $LaIn_3$ [3], and is the principal experimental result of this work. Two features of Fig. 4 give insight into the diffusion mechanisms: (a) The mean jump frequency generally increases with $n$; (b) The jump frequency activation enthalpy is the same for $n=3$, 5 and $\infty$. A simple model is proposed to explain both features. As described above, lack of nuclear relaxation of the site 4 signal indicates that there is negligible diffusion of probe atoms across the cobalt layers (jump rates are less than the measurement threshold of ~1 MHz). The activation enthalpy for jumps in general can be considered to be the sum of an effective enthalpy for formation of vacancies on the In-sublattice and a migration enthalpy for probe atoms to jump into an adjacent vacancy. Since the activation enthalpy for jumps of probe atoms on sites-1 is the same for $n=3,5$ and $\infty$, we assume that the formation and migration enthalpies, and diffusion mechanism are independent of $n$.

What differs with $n$ is the average connectivity of near-neighbor jumps of atoms on type-1 indium sites to other type-1 indium sites. Consider the $n=2$ structure in Fig. 1. There are two type-1 atoms in the mixed La-In layers per unit cell, each of which has four type-1 near-neighbors. The other two type-1 atoms are in the median layer, with each having 8 type-1 near neighbors. Therefore, the average connectivity of probes on type-1 sites to other type-one sites is 6. Analogous considerations show that in general the average connectivity of site-1 atoms as a function of $n$ is $24(n-1)/(3n-2)$, which gives the expected limits 0 for $n=1$ (since the type-1 atom has no near-neighbor type-1 sites) and 8 for $n=\infty$. Fig. 5 shows how a plot of jump frequencies at 700 K (right-hand scale) correlates with the site connectivity (left-hand scale), both graphed as a function of $1/n$.

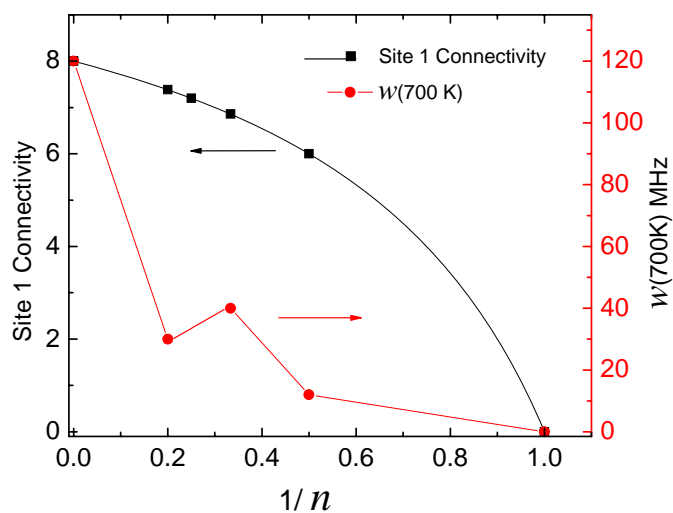

Fig. 5. Overlay of plots of jump frequencies at 700 K (right-hand scale) and the average diffusion connectivity of atoms on sites-1 (left-hand scale), both graphed as a function of $1/n$.

Examination of Fig. 5 shows that consideration of the average site-connectivity for diffusion in layers of site-1 atoms gives a qualitative explanation for the dependence of jump frequencies on $n$.

**Summary and acknowledgment**

Frequencies of jumps of Cd probe atoms on In sites were measured in $La_nCoIn_{3n+2}$ phases having compositions close to $n$= 1, 2, 3 and 5. Lack of nuclear relaxation for probe atoms on In-sites next to the Co-planes (site 4) indicates that there is negligible diffusion across the Co-planes. Jump frequencies were found to be less than 1 MHz for $n$=1 but to increase with $n$. Jump-frequency activation enthalpies for $n$=3 and 5 were observed to be the same as for $n=\infty$ ($LaIn_3$), suggesting the same type of diffusion mechanism. The increase in jump rate as $n$ is shown to be expected when one takes into account the increasing connectivity of the indium diffusion sublattice as the density of Co-planes decreases.

This work was supported in part by the National Science Foundation under grant DMR 09-04096 (Metals Program)

**References**

[1] M.O. Zacate, A. Favrot and G.S. Collins, Phys. Rev. Lett. 92, (2004) 22590.

[2] Gary S. Collins et al., Hyperfine Interactions 159 (2005) p. 1.

[3] G.S. Collins et al., Defect and Diffusion Forum 237-240 (2005) p. 195.

[4] Gary S. Collins et al., Phys. Rev. Lett. 102 (2009) p. 155901.

[5] Tyler Park et al., Hyperfine Interact., DOI 10.1007/s10751-011-0332-6 (online 15 Apr 2011)

[6] Robin T. Macaluso, et al., Journal of Solid State Chemistry 166 (2002) p. 245.

[7] H. Hegger, et al., Phys. Rev. Lett. 84 (2001) p. 4986.

[8] J.D. Thompson, et al., Physica B 329-333 (2003) p.446.

[9] Günter Schatz and Alois Weidinger, *Nuclear Condensed Matter Physics* (Wiley, 1995).